# A New Strategy for the Morphological and Colorimetric Recognition of Erythrocytes for the Diagnosis of Forms of Anemia based on Microscopic Color Images of Blood Smears

J. Nango ALICO[1], Sié OUATTARA[2]*

URMI Electronique et Electricité Appliquées (EAA)
Laboratoire des Sciences et Technologies de la
Communication et de l'Information (LSTCI)
Institut National Polytechnique Felix Houphouët-Boigny
(INP-HB), BP 1093 Yamoussoukro, Côte d'Ivoire

Alain CLEMENT[3]

Laboratoire Angevin de Recherche en Ingénierie des Systèmes
(LARIS), Institut Universitaire de Technologie (IUT) /
Université d'Angers, 4 Boulevard Lavoisier - BP 42018 -
49016 - ANGERS, France

*Abstract*—The detection of red blood cells based on morphology and colorimetric appearance is very important in improving hematology diagnostics. There are automatons capable of detecting certain forms, but these have limitations with regard to the formal identification of red blood cells because they consider certain cells to be red blood cells when they are not and vice versa. Other automata have limitations in their operation because they do not cover a sufficient area of the blood smear. In spite of their performance, biologists have very often resorted to the manual analysis of blood smears under an optical microscope for a morphological and colorimetric study. In this paper, we present a new strategy for semi-automatic identification of red blood cells based on their isolation, their automatic color segmentation using Otsu's algorithm and their morphology. The algorithms of our method have been implemented in the programming environment of the scientific software MATLAB resulting in an artificial intelligence application. The application, once launched, allows the biologist to select a region of interest containing the erythrocyte to be characterized, then a set of attributes are computed extracted from this target red blood cell. These attributes include compactness, perimeter, area, morphology, white and red proportions of the erythrocyte, etc. The types of anemia treated in this work concern the iron-deficiency, sickle-cell or falciform, thalassemia, hemolytic, etc. forms. The results obtained are excellent because they highlight different forms of anemia contracted in a patient.

## I. INTRODUCTION

Blood is a complex connective tissue that is subject to constant renewal. It is made up of a set of elements suspended in a liquid. These various elements, which are cells each performs specific functions necessary for the maintenance of life [1]. We can essentially distinguish three categories of circulating cells of various shapes, sizes and colors:

- Red blood cells or erythrocytes come from bone marrow and are essentially made up of hemoglobin,
- White blood cells or leukocytes and
- Platelets or thrombocytes.

The specific role of red blood cells is to favor gas exchange between deep cells through the lungs. White blood cells defend the body against aggression by pathogens and toxins, while platelets favor blood clotting in the event of injury by forming a platelet clot instead of being lost [1]. Healthy red blood cells have regularly rounded shapes (circular shape) on a blood smear with a slightly clear central area. This area is due to the absence of a nucleus. Red blood cells are practically uniform in size, the red color of a healthy red blood cell occupies most of the cell. These characteristics are altered by the presence of microbes, by the insufficiency or lack of certain nutrients in the body. The deformation and alteration of the color of the red blood cells is regularly indicative of a pathology. So, the recognition of each of these forms will allow us to have appropriate information on the related disease. The analysis and manually recognition of blood cells (white and red blood cells) from blood smears is a slow, tedious and subjective task. In addition, it depends on the skills and experience of the technician and can often lead to repeatability [2, 3, 4] of the tests. Therefore, the implementation of an automatic and accurate diagnostic procedure is necessary. For this purpose, there are automatons capable of detecting certain forms of red blood cells, but these have limitations in terms of their formal identification as they sometimes consider in practice a cluster of red blood cells to be a single red blood cell or often a large cluster is not considered to be a red blood cell, as shown by Godon A et al. in 2012 through their work on errors in blood counts produced by automatons [5]. In the thesis work presented by GRIMON Noemie in 2019 on the evaluation of automated microscopes, he showed that despite the speed and performance of automated microscopes, they present insufficiencies on the reading surface of blood smears, and revealed that, despite everything, digital cyto-morphology is still dependent on the visual examination of biologists [6]. Although time saving is an important factor in medical diagnosis, the reliability of the results remains the major element in the establishment of an effective treatment protocol. With this in mind, we propose a new strategy for the morphological and colorimetric identification of erythrocytes for the diagnosis of forms of anemia based on microscopic color images of Blood Smear. Following this approach, many researchers have tried to automate the detection, recognition and classification of blood cells from microscopic blood smear images using an image processing approach to help hematologists diagnose many diseases quickly and accurately. R. Tomaria et al. 2014 applied the morphological parameter compactness and the seven HU moment characteristics for red blood cell detection. The authors also developed an algorithm based on artificial neural networks to assign each cell the corresponding class [7]. In 2017 H.A. Elsalamony, for an identification of the normal blood cells, sickle cell and elliptocyte cells proceeded by a signature of each form by extracting the different morphological parameters characterizing each of the forms [8]. M. M. Abdul Jamil et al in 2019 used a marker-controlled watershed segmentation method to solve the thorny problem of cell overlap to detect blood cells. But this method increases the complexity of the algorithm [9]. The detection of blood cells, namely red and white blood cells, was the subject of work by C. Di Ruberto et al. in 2019. In their work, the authors used an Edge Boxes approach for simple cell detection [10]. Although efficient, this method does not allow us to appreciate certain morphologies and colorimetric aspects of the cells which are the basis of the importance of our work.

*Corresponding Author

This research paper is organized as follows: Section II allows us to review the work carried out in the literature, Section III will deal with the materials and methods, the results will be discussed in Section IV and finally we will conclude with a conclusion and perspectives in Section V.

## II. RELATED WORK

Manual analysis and recognition of blood cells (white blood cells and red blood cells) from blood smears is a slow, tedious and subjective task. It also depends on the skills and experience of the technician and the results of analyzes show repeatability [2, 3, 4]. Therefore, an automatic and precise mechanism is needed to deal with the above-mentioned problems.

Many researchers have tried to automate the detection, recognition and classification of blood cells from microscopic images of blood smears using an image processing approach to help hematologists diagnose many diseases in a meaningful way fast and precise.

Given the importance of the subject, several studies have proposed methods for the automatic recognition and classifications of blood cells based on shape descriptors extracted from them [11, 12, 13]. Chettaoui. C. et al [14,] used several descriptors to identify the sickle red blood cell as in the work of. In this article the authors tested the correctness of each descriptor, namely the descriptors of the extremities, of the circumscribed inscribed circle, of Fourier and of Fourier Mellin. After the test carried out on each one, they opted for the descriptor of the inscribed-circumscribed circle as being the most precise.

In [15], the authors identified certain anemic red blood cells based on the automatic processing of images acquired on blood smears taken. Initially, they used the gray world technique and the geometric mean filter to respectively reduce the backlighting of the different smears and the noise associated with the images during the acquisition. Secondly, the cells were segmented using the watershed segmentation technique controlled by marker. For a better identification the authors used the logistic regression classiffer with a high precision of the order of 86.87% compared to the standard classiffers. These classifiers can be grouped into three (3) categories: supervised classifiers (SVM, Bayes algorithms, K-NN, Neural networks, CNN, Multilayer Perceptron, decision tree, etc.) [1- 16], non-classifiers -supervised (K-means, C-means, Fischer) [17] and hybrid classifiers [17].

Recently, D. A. TYAS et al [18], combined 64 characteristics to identify nine forms of red blood cells. This combination is based on the extraction of three types of characteristics: morphological characteristics, color and texture. The extraction of the different characteristics was carried out using the gray level co-occurrence matrices for the texture characteristics, the use of the color moments for the different color parameters, the moment invariants and the geometric parameters were proposed by the authors for the extraction of morphological parameters. Following this first step, the authors proposed the machine learning algorithms, namely the multilayer perceptron (MLP) with learning methods by back propagation for the classification of red blood cells.

## III. MATERIALS AND METHODS

### A. Materials used

This work to identify forms of anemia in patients required blood samples. These samples helped to the preparation of blood smears, i.e. five (05) smears per patient, two (02) of which are retained per patient after physical analysis. The smears were then digitized through our images acquisition system composed of a microscope brand MOTIC equipped with a 6V-220W halogen lamp for the illumination of the sample coupled with a colour CMOS camera brand Moticam2 with a spatial resolution of 2 Megapixels. The latter is connected to a computer through a USB 3.0 port for the acquisition of digital microscopic colour images.

### B. Methods

A medical application requires a precision in the identification in the diagnosis in order to propose the most appropriate clinical treatment because human life is delicate. In view of recent work [7, 8, 9, 10], a purely automatic processing and analysis of digital images induces errors in the detection of red blood cells, their morphologies and others. Therefore, in order to have a precise diagnosis, we propose a semi-automatic approach which consists in choosing in a first step, regions of interest, each containing red blood cells to be isolated, processed and analysed in order to solve upstream problems such as overlapping red blood cells, open red blood cells, etc.

This work was carried out in collaboration with the Yamoussoukro Regional Hospital Centre (CHR) and the Yamoussoukro Blood Transfusion Centre (CTS). Blood samples were taken from target patients known to the services of the CHR and the CTS concerning patients suffering from certain forms of anaemia. These are the ferriprive, sickle-cell or falciform, thalassemic, hemolytic forms, etc.

The blood smears were performed on different samples taken from the target patients. For the acquisition of the microscopic colour images from the blood smears, an adequate configuration of our instrumentation presented in Fig. 1 in the previous section was necessary to obtain well-contrasting and sharp images.

The following sections present the digital image acquisition protocol and the proposed semi-automated diagnostic method for the identification of the forms of anaemia a patient suffers from.

It should be noted that the blood smear is an examination prepared on a thin, stained slide and its analysis allows a morphological and colorimetric study of blood cells, in this case red blood cells or erythrocytes [6, 19].

*1) Image acquisition protocol:* The acquisition of the different images required the appropriate configuration of the camera and the microscope. The resolution of the images is 1600 x 1200. The colour's gains in the different channels are as follows: Red at 2.04, Green at 1.2, Blue at 1.12 when the white balance is activated. This configuration of the camera associated with that of the microscope as well as the illumination of the samples allowed us to acquire a series of images. The acquisition process is triggered as soon as the operator has a blood smear of quality duly prepared according to the procedure recommended by the World Health Organization (WHO) [20]. The blood smear is placed on the microscope stage as soon as the acquisition software is launched and then several images are acquired by connecting the camera to the computer using at least the USB cable. The optical images thus obtained were obtained with the objective 100 times (100X), which is equivalent to a magnification (G) of 1000 (eyepiece: 10X and objective 100X, i.e. a G = 10 X 100), giving a clear and well-contrasting morphological and colorimetric characterization of the red blood cells.

This step led to a first contribution which is the implementation of a database of microscopic colour images of patients.

We present some image acquisition results illustrated in Fig. 1.

The image in Fig. 1(a) shows red blood cells of regularly rounded shapes and uniform colouring with a slightly light-coloured central area: these are the characteristics of normal-looking red blood cells. Meanwhile, the images in Fig. 1(b) shows pale red blood cells with a largely light central area and a red coloration located at the periphery forming a ring. These are diseased red blood cells characteristic of a specific form of anaemia. In addition, while the image in Fig. 1(c) shows elongated elliptical red blood cells.

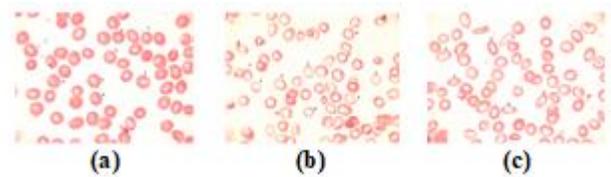

Fig. 1. Healthy Red Blood Cells, (b) Ring-Shaped Red Blood Cells (Annulocytes), (c) Elongated Red Blood Cells (Ovalocytes or Elliptocytes).

*2) Methods for the identification of the proposed forms of anaemia.*

*a) Principle of the identification method:* After the acquisition of microscopic color images of blood smears, we developed an application whose principle is illustrated in the flowchart in Fig. 3.

Its principle consists first of choosing an image of a patient's blood smear, then the application allows the selection of an area of interest containing the target red blood cell to be treated.

This area is segmented using a modified version of Otsu's segmentation algorithm combined with a search for related components in eight connexities under certain constraints to isolate the red blood cell followed by a partition of the red blood cell into two classes. The first class is the red area of the red blood cell and the second is the white area.

After the previous step of isolation and segmentation of the red blood cell, we calculate different morphological and colorimetric quantities of the red blood cell. These include, among others, the following: (i) the areas occupied by the red and white zone of the red blood cells, (ii) the total surface area of the red blood cell, (iii) the circumference of the red blood cell, (iv) the compactness, (v) the shape of the red blood cell, etc.

Finally, a test on the morphology and color of the red blood cell is carried out using the previously calculated quantities. From this step it is possible to detect a healthy or sick red blood cell. For a diseased red blood cell, it can be a question of identifying red blood cells such as sickle-shaped, circular (annulocytes), elliptical, pear-shaped, macrocytes, microcytes, hypochromic red blood cells, acanthocytes, etc. or combinations of shapes.

*b) Semi-supervised selection of a region of interest and isolation of the target red blood cell:* Following image acquisition, the proposed method requires the intervention of the operator, in this case the biologist. He selects through the developed application the region of interest containing the red blood cell to be identified. It is on this region that we apply Otsu's binary segmentation algorithm, combined with a search for related components under certain constraints that allows us to isolate the target red blood cell. The principle of related components is presented in the following sections. The binarization method by automatic thresholding makes it possible to separate the object from the background of the image, its principle is presented in the following section. An example of the isolation of the target hematite from the region of interest by our application is shown in Fig. 2.

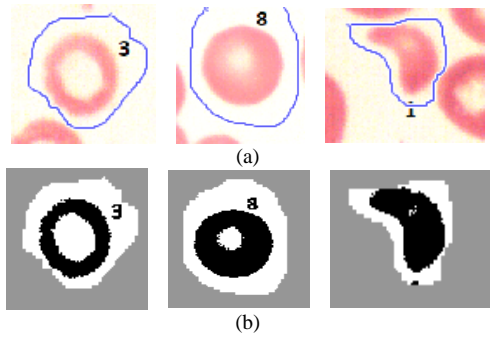

Fig. 2. (a) Region of Interest in Source Image, (b) Target Cell to be Isolated.

- Principle of the Otsu binary segmentation algorithm.

Let x be a variable describing the greyscale values of the image.

We wish to construct two (02) classes C1 and C2. Let P1(x) and P2(x) be the probabilities of classes C1 and C2.

Let σ1 and σ2 be the respective intra-class variances of classes C1 and C2.

Otsu's principle consists in choosing the threshold x minimizing the total intra-class variance [21, 22] that is minimiser σ_Total (x). Let:

$$\sigma_{Total}(x) = \mu_1(x)\sigma_1^2(x) + \mu_2(x)\sigma_2^2(x) \qquad (1)$$

Or to maximize inter-class energy i.e. maximize σ inter (x). Let it be:

$$\sigma_{inter}(x) = \sigma^2 - \sigma_{Total}^2(x) = P_1(x)P_2(x)[\mu_1(x) - \mu_2(x)]^2 \qquad (2)$$

The description of the algorithm is summarized in three steps as shown in the flowchart in Fig. 4.

- Related components and isolation of the red blood cell

After segmentation of the selected area by Otsu's method following automatic threshold search, it was possible to have a binarized image in which the region containing the red blood cell is the large black area as shown in Fig. 2(b), and thus an isolation of the target red blood cell was achieved. This isolation of red blood cells was made possible by the implementation of a related component search algorithm in eight (08) connects (see Table I for the related neighbours of the pixel (x, y)). Some examples of red blood cell isolation are shown in Fig. 5.

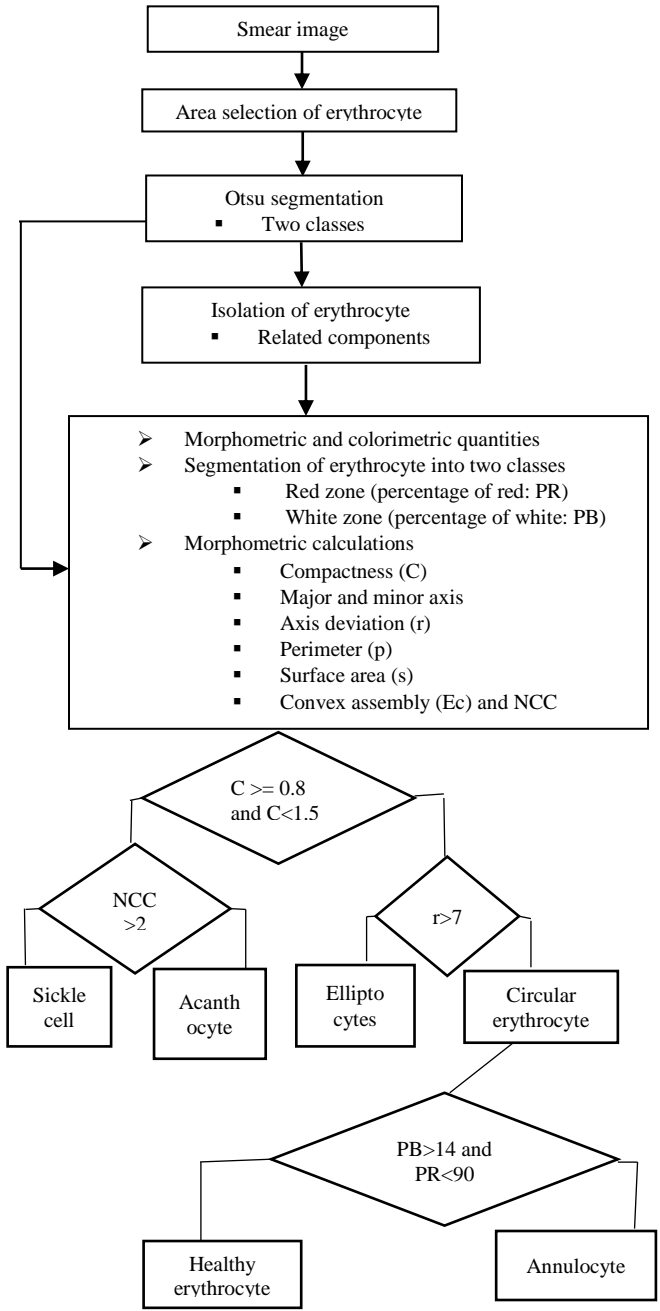

Fig. 3. Flowchart of the Method for the Identification of Erythrocytes Forms.

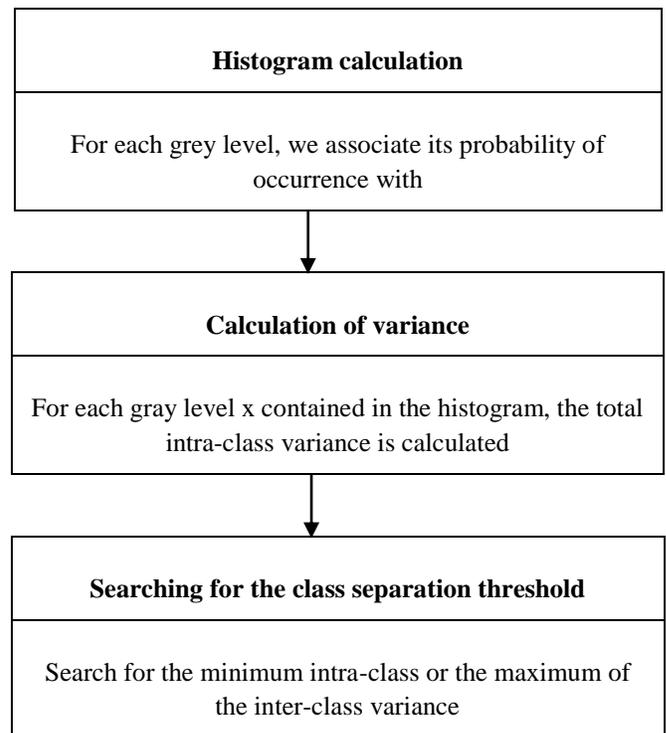

Fig. 4. Description of the Algorithm Steps.

TABLE I. TABLE SHOWING THE EIGHT CONNECTIONS OF A PIXEL (X, Y)

| (x-1, y-1) | (x, y-1) | (x+1, y-1) |
|---|---|---|
| (x-1, y) | **(x, y)** | (x+1, y) |
| (x-1, y+1) | (x, y+1) | (x+1, y+1) |

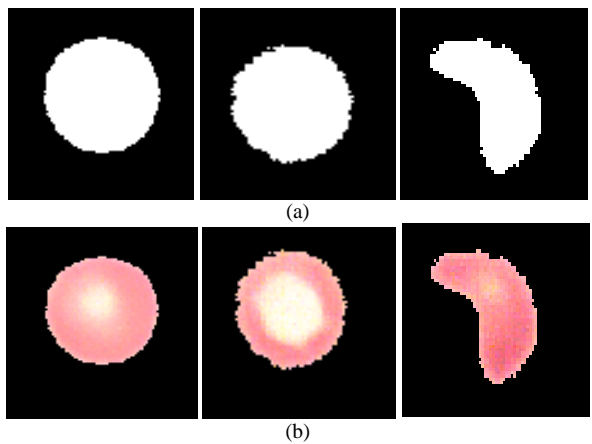

(a)

(b)

Fig. 5. (a) Erythrocyte Highlighted in the Segmentation Zone. (b) Erythrocyte Isolated from the Selection Area in the Real Image.

*c)* Estimation of morphometric and colorimetric magnitudes and implementation of a shape-testing algorithm for the red blood cells.

The calculation of the different morphometric and colorimetric quantities in a patient requires the complete isolation of the target red blood cells.

For the morphometric quantities, the estimation of parameters such as the area, perimeter, compactness of the red blood cells and others was discussed. The direct use of these morphometric quantities does not make it possible to directly predict the shape of a target red blood cell. Therefore, we have developed a clever algorithm to identify the shape of a hematite based on other quantities such as the major axis, minor axis, convex set, etc. These quantities are presented in the following sections.

As far as colorimetric quantities are concerned, we have estimated the average colour of each target hematite, the average colours of the red and white areas of the hematite as well as their proportions and others. Their calculations are also presented in a section below.

In sum, the calculation of the morphometric and colorimetric quantities was possible thanks to algorithms that we implemented in the MATLAB programming environment.

- Morphometric parameters

The pattern recognition system is undeniably based on different parameters. These parameters are measurements extracted from the object shapes [23]. Healthy red blood cells generally have circular shapes and a pinkish coloration on an MGG smear. All changes in each of these aspects are due to haematological pathologies such as anaemia. Anaemia is a disease related to red blood cells. It can be detected on a blood smear based on the staining and morphology of the red blood cells. The characterization of each form is made possible thanks to the morphometric parameters extracted from each healthy or diseased red blood cell.

The following Table II presents some parameters with their different formulas.

TABLE II. GROUPING THE EQUATIONS (EQUITATION 3 TO EQUATION 11) AND THEIR DESCRIPTION

| Parameter | Equations | Descriptions |
|---|---|---|
| Area | $area = \sum_x \sum_y f(x,y)$ **(3)** | Set of pixels covering the region that the cell represents. |
| Perimeter (**P**) | $P = \sum_x \sum_y f(x,y)$ such as $x, y \in F(R)$ **(4)** | The total number of pixels along the edge of the cell. |
| Compactness | $C = \frac{4\pi.aire}{P^2}$ **(5)** | It is a morphological quantity that participates in the characterization of certain object shapes. |
| Major axis | **Majaxis**=max {dst (a, b) │a, b∈R} **(6)** <br> **Majaxis**=max {dst (a, b) │a, b∈F(R)} **(7)** | These axes are obtained after the determination of the spatial barycentre of the isolated and binarized red blood cell. |
| Minor axis | **Minaxis**=min {dst (a, b) │a, b∈R} **(8)** <br> **Minaxis**=min {dst (a, b) │a, b∈F(R)} **(9)** | The minor axis (Minaxis) is the smallest diameter of the cell passing through the barycentre. |
| Axis spacing | **Axis spacing** = Majaxis - Minaxis **(10)** | The difference between the axes allows us to make a clear difference between certain forms of erythrocyte. |
| Convex set | {(C) is convex if ∀x, y∈C ∀ t ∈ [0,1] tx+ (1-t)y ∈ C} **(11)** | A set (C) is said to be convex when any point x and y of the set describes a segment that is entirely part of the set. |

- Colorimetric parameters

On a blood smear, red blood cells appear pinkish with a slight whitish area in the centre. A change in this appearance is usually indicative of a pathology. The proposed algorithm allows us to quantify all the pixels belonging to each area (pink or white). Indeed, the transformation of the colour image into a grey-level image followed by the application of Otsu's method combined with a two-class segmentation on the isolated cell, allowed us to highlight the two colours that make up the red cell (red represented by black and white) and to quantify them through the calculation of the image histogram. The histogram is a statistical graph that indicates the number of pixels per grey level.

The pixels thus grouped according to their grey level give rise to two regions R1 (red) and R2 (white) in healthy erythrocytes and in some binarized diseased erythrocytes. The ratio of these two quantified regions is an important piece of information in the formal characterization of certain anaemic forms. These two quantities are calculated automatically by the proposed method following the relationships below:

$$NbrePixelWhite = \sum_x \sum_y f(x,y) \text{ such that } I(x,y) = 255 \quad (12)$$

$$NbrePixelRed = \sum_x \sum_y f(x,y) \text{ such that } I(x,y) = 0 \quad (13)$$

NbrePixelsWhite and $NbrePixelRed$ indicate the pixel size of the white region and the pixel size of the red region of the binarized erythrocyte, respectively. In this binarization,

I(x, y) is the gray level value of the above-mentioned regions.

The determination of the proportion of distribution of pixels in each of the regions is obtained from the following equations:

$$PR = \frac{1}{aire} \sum_x \sum_y f(x,y) \text{ such that } (x,y) \in \textbf{R1} \quad (14)$$

$$PB = \frac{1}{aire} \sum_x \sum_y f(x,y) \text{ such that } (x,y) \in \textbf{R2} \quad (15)$$

PR is the percentage of distribution of pixels in the red region (R1) and PB is the percentage distribution of pixels in the white region (R2) in the binarized erythrocyte.

- Algorithm for testing the morphology of red blood cells

This section presents the subtlety developed for the detection of red blood cell geometry. The search for the morphology of red blood cells requires first of all knowledge of the compactness which makes it possible to differentiate in a first step a circular or elliptical shape from other shapes such as falciforms and acanthocytes. After learning, we find that the latter have a compactness of less than 0.8.

In a second step:

- If the compactness is greater than or equal to 0.8, we try to find out whether the red blood cell is circular or elliptical; under these conditions, we calculate the major and minor axes of the red blood cell, then we deduce the difference *r* between these two axes. If *r* is greater than 7 pixels then its shape is elliptical, otherwise it is circular.

- Otherwise, i.e. if the compactness is less than 0.8 then it is non-convex or concave. In this case, we calculate the barycentre of the hematite and then we look for the point furthest from it. Then, a square mask of variable side limited by the farthest point of the erythrocytes is generated. From the erythrocytes, we isolate the complementary of the intersection of the erythrocytes with the mask. This region is labelled in related components. When the number of components is greater than two, it is an acanthocyte form; conversely, if this number is equal to two, we detect a falciform form.

## IV. RESULTS AND DISCUSSIONS

The method of identifying the forms of red blood cell disease proposed through the flow chart in Fig. 3 allowed us to estimate the different morphological and colorimetric quantities from blood smear images obtained from different patients, both healthy and anaemic. These quantities allow us to clearly identify the different red blood cells, both healthy and diseased. Thus, this diagnosis will enable pathologists to propose appropriate treatments. Our samples come from anaemic and non-anaemic patients known through a blood count carried out by the existing automatons. Our method has made it possible to correct the insufficiencies of the automatons to effectively detect certain forms of anemia and certain pathologies linked to the morphological aspect and the color of the red blood cells, in particular sickle cell anemia, anemia due to alcoholic cirrhosis, hereditary elliptocytosis, etc. In this section, we present some of the results obtained. As well as their discussion on patients identified as healthy and anaemic by the count examination. It should be noted that the CBC scan does not indicate the precise form of anemia in the patient.

### A. Analysis of the Results of our Application on a Healthy Patient

Following the selection of the target red blood cells through our application, Fig. 6 presents the steps of red blood cell isolation in a known healthy patient. In addition, Tables III and IV present the colorimetric and morphometric parameters, as well as the test of the geometry of the selected red blood cells which lead to the identification of the form of anemia or non-anemia of this patient.

From the colorimetric analysis, it appears from Table III that in a healthy red blood cell the pixels designating the red color indicating the presence of hemoglobin occupy 86 to 90% of the surface area of the red blood cell and the clear central area represents only 10 to 14% of the cell area. This finding is consistent with data in the literature. Morphologically, to show its circularity, we calculated the compactness of the cell and the distance r between the major and minor axes of the cell. For a healthy red blood cell that has a circular shape, the distance r between the major and minor axes is very small, i.e. less than 7 pixels, and the compactness is close to the value 1. This confirms the data in Table IV where the variable distance r varies between 2 and 6 with compactness values between 1.1 and 1.2. The normal colorimetric proportion in white and red pixels added to the estimation of the two previous parameters allow us to identify with certainty that the red blood cells of this patient are healthy.

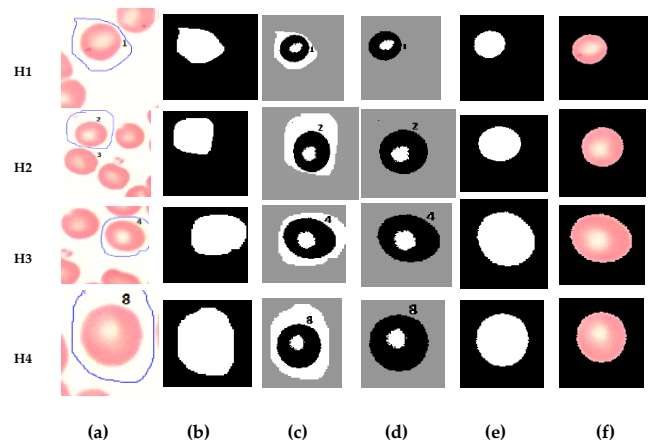

Fig. 6. (a) Region of Interest in the Source Image (b) Selection Area Binarized by Otsu's Algorithm (c) Erythrocyte Isolation (d) Segmentation into Two Classes (e) Isolated Segmented Erythrocyte (f) Isolated Source Erythrocyte.

TABLE III. COLORIMETRIC PARAMETERS OF HEALTHY RED BLOOD CELLS

| | Class size | | Color | | | Area | Distribution | |
|---|---|---|---|---|---|---|---|---|
| | Red Pixels | White Pixels | Red | Green | Blue | In pixel | % White | % Red |
| H1 | 3889 | 583 | 255 | 222 | 219 | 4472 | 13.03 | 86.94 |
| H2 | 4321 | 618 | 255 | 222 | 219 | 4939 | 12.51 | 87.49 |
| H3 | 4054 | 643 | 255 | 222 | 219 | 4697 | 13.69 | 86.31 |
| H4 | 4253 | 469 | 255 | 222 | 219 | 4722 | 10.00 | 90.00 |

H1, H2, H3 and H4 represent red blood cells.

TABLE IV. MORPHOLOGICAL PARAMETERS OF HEALTHY RED BLOOD CELLS

| | Compac | Area | Peri | MinorA | MajorA | Spacing | Varconvex |
|---|---|---|---|---|---|---|---|
| $H_1$ | 1,22 | 4472 | 215 | 35.36 | 39.22 | 3.86 | 0 |
| $H_2$ | 1,20 | 4939 | 231 | 37.40 | 41.04 | 3.63 | 0 |
| $H_3$ | 1,13 | 4697 | 228 | 35.33 | 41.74 | 6.41 | 0 |
| $H_4$ | 1,20 | 4722 | 224 | 37.02 | 39.73 | 2.71 | 0 |

Compact: compactness, Peri: Perimeter, MinorA: minor axis, MajorA: major axis, Spacing: difference between majoA and MinorA.

*1) The results of our application on four anaemic patients:* We present the diagnostic results of the forms of anemia of four (04) anaemic patients produced by our application. For clarity, we have named the patients from Patient 1 to Patient 4.

*a) Analysis of Patient 1 results:* The results from our application for Patient 1 shown in Fig. 7 and in Table VI show that the compactness of the red blood cells is between 0.9 and 1, confirming their circular shape. Moreover, the value of the deviation r of the axes below 7 pixels confirms this shape. Consequently, the red blood cells have a normal morphology.

The colorimetric characteristics of the red blood cells grouped in Table V indicate that the proportion in white pixels varies from 33.26% to 45.5% and the proportion in red varies from 54% to 66%. The maximum proportion, which is 66.79%, is below the standard for a healthy red blood cell, whose proportion varies between 86% and 90%. The observation made on the colorimetric characteristics is the best indicated for the identification of this form of red blood cell called annulocyte generally indicating a martial deficiency anemia.

*b) Presentation of patient 2 results:* Sickle cell anemia is a hereditary and anaemic disease that cannot be detected by the blood test performed by current automatic machines. This serious and painful disease deserves early management following a clear and precise diagnosis. Our proposed method fulfils these conditions because the examination of the blood smear facilitates the morphological study of the red blood cells. Since sickle cell disease transforms red blood cells into a particular form: sickle or lunar crescent, its detection on a smear is easier.

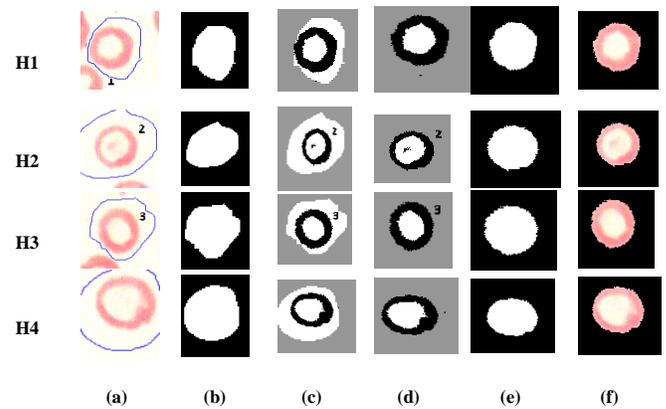

Fig. 7. (a) Region of Interest in the Source Image (b) Selection Area Binarized by Otsu's Algorithm (c) Erythrocyte Isolation (d) Segmentation into Two Classes (e) Isolated Segmented Erythrocyte (f) Isolated Source Erythrocyte.

TABLE V. COLORIMETRIC PARAMETERS FOR ANNULOCYTE CHARACTERIZATION

| | Class size | | Color | | | Area | Distribution | |
|---|---|---|---|---|---|---|---|---|
| | Red Pixels | White Pixels | Red | Green | Blue | In pixel | % White | % Red |
| $H_1$ | 2385 | 1186 | 255 | 232 | 221 | 3571 | 33.26 | 66.79 |
| $H_2$ | 1913 | 1597 | 255 | 232 | 221 | 3510 | 45.50 | 54.50 |
| $H_3$ | 2341 | 1248 | 255 | 232 | 221 | 3589 | 34.77 | 65.23 |
| $H_4$ | 2021 | 1525 | 255 | 232 | 221 | 3546 | 43.00 | 57.00 |

TABLE VI. MORPHOLOGICAL PARAMETERS FOR ANNULOCYTE CHARACTERIZATION

| | Compac | Area | Peri | MinoA | MajorA | Spacing | Varconvex |
|---|---|---|---|---|---|---|---|
| $H_1$ | 0,93 | 3571 | 219 | 30.63 | 35.27 | 4.64 | 0 |
| $H_2$ | 1,05 | 3510 | 205 | 30.87 | 34.82 | 3.95 | 0 |
| $H_3$ | 0,91 | 3589 | 222 | 29.92 | 36.44 | 6.52 | 0 |
| $H_4$ | 1,04 | 3546 | 207 | 29.51 | 36.14 | 6.63 | 0 |

The colorimetric characteristics of red blood cells in Fig. 8 and grouped in Table VII show that the proportion of red pixels varies between 91% and 99%, which suggests the proportion of a healthy red blood cell. On the other hand, the morphological data in Table VIII indicate that the compactness, which varies from 0.5 to 0.7, shows us that these red blood cells are quite different from normal red blood cells whose compactness varies from 0.9 to 1 as shown in Table IV. Moreover, the difference r between the different axes confirms the non-normality of this form of red blood cell, since this difference r is around 35 pixels, which is much greater than the normal, which is less than 7 pixels. Also given the shape of the red blood cells in Fig. 8, we used the convex set principle for a formal identification of them. This is confirmed by the value of the varconvex variable in Table VIII, which takes the value 1 when it is a non-convex shape corresponding to the sickle cell morphology. Next, a square mask with a variable side bounded by the farthest point of the red blood cell. From the red blood cell, we isolate the complementary of the intersection of the red blood cell with the mask. This region is labelled in related components. When the number of components is equal to two as shown by the value of the variable NCC (number of related components) which is exactly 2, it is a sickle shape. This confirms the visual appearance of these red blood cells. Our method thus makes it possible to clearly identify the sickle cell red blood cells in order to institute a treatment protocol adapted for this type of anemia.

*c) Presentation of patient 3 results:* After the results obtained and indicated in Fig. 9, the different steps leading to the isolation of the selected red blood cell, we present in Tables IX and X the different colorimetric and morphological characteristics.

The analysis of the colorimetric parameters in Table IX shows a total absence of white pixels and the presence of 100% of the proportion of red pixels, which indicates an anomaly in the distribution of white and red pixels in a healthy red blood cell. Indeed, in a healthy red blood cell the proportion of white pixels varies between 10% and 14% of the total area of the red blood cell and the proportion of red pixels varies between 86% and 90% as shown in Table III.

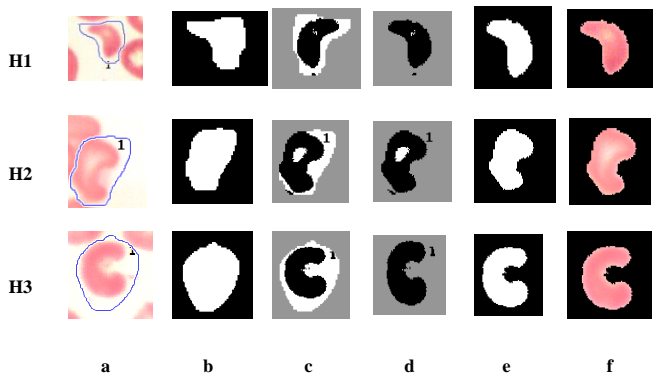

a b c d e f

Fig. 8. (a) Region of Interest in the Source Image (b) Selection Area Binarized by Otsu's Algorithm (c) Erythrocyte Isolation (d) Segmentation into Two Classes (e) Isolated Segmented Erythrocyte (f) Isolated Source Erythrocyte.

TABLE VII. SICKLE-CELL COLORIMETRIC PARAMETER

| | Class size | | Color | | | Area | Distribution | |
|---|---|---|---|---|---|---|---|---|
| | Red Pixels | White Pixels | Red | Green | Blue | In pixel | % White | % Red |
| H$_1$ | 2022 | 15 | 253 | 214 | 204 | 2037 | 0.73 | 99.27 |
| H$_2$ | 3547 | 291 | 255 | 213 | 206 | 3838 | 08.20 | 91.8 |
| H$_3$ | 3789 | 2 | 255 | 218 | 215 | 3791 | 0.05 | 99.95 |

TABLE VIII. SICKLE-CELL MORPHOLOGICAL PARAMETERS

| | Compac | Area | Peri | MinorA | MajorA | Spacing | varconvex | NCC |
|---|---|---|---|---|---|---|---|---|
| H$_1$ | 0.67 | 2037 | 195 | 8.49 | 40.49 | 32.00 | 1 | 2 |
| H$_2$ | 0.76 | 3838 | 252 | 12.21 | 47.40 | 35.19 | 1 | 2 |
| H$_3$ | 0.53 | 3791 | 300 | 10.24 | 45.93 | 35.69 | 1 | 2 |

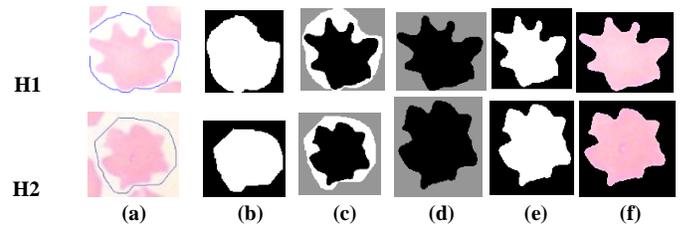

(a) (b) (c) (d) (e) (f)

Fig. 9. (a) Region of Interest in the Source Image (b) Selection Area Binarized by Otsu's Algorithm (c) Erythrocyte Isolation (d) Segmentation into Two Classes (e) Isolated Segmented Erythrocyte (f) Isolated Source Erythrocyte.

TABLE IX. COLORIMETRIC PARAMETERS OF ACANTHOCYTES

| | class size | | Color | | | Area | Distribution | |
|---|---|---|---|---|---|---|---|---|
| | Red Pixels | White Pixels | Red | Green | Blue | In pixel | % Blanc | % Rouge |
| H$_1$ | 6489 | 0 | 254 | 222 | 229 | 6489 | 0 | 100 |
| H$_2$ | 5785 | 0 | 254 | 222 | 229 | 5785 | 0 | 100 |

TABLE X. MORPHOLOGICAL PARAMETERS OF ACANTHOCYTES

| | Compac | Area | Peri | MinoA | MajoA | Spacing | varconvex | NCC |
|---|---|---|---|---|---|---|---|---|
| H$_1$ | 0.57 | 6954 | 378 | 24.41 | 66.43 | 42.02 | 1 | 4 |
| H$_2$ | 0,78 | 5785 | 306 | 30.30 | 51.58 | 21.28 | 1 | 6 |

At the morphological level, the analysis of Table X gives a compactness value of 0.5 to 0.7. This variation is almost identical to the compactness of the sickle shapes as shown in Table VIII. This confusion is also observed at the level of the deviation r of the major and minor axes of the erythrocyte which is between 21 pixels and 41 pixels. So, for an accurate identification of this form of red blood cell, we used our red blood cell morphology testing algorithm described in section above. The value of the variable NCC which indicates the number of related components obtained by labelling the areas in related components following the intersection of the cell with the mask. In our coding, when the number of related components is greater than two (02) then we detect an acanthocyte which is the form of red blood cell disease observed in hepatic cirrhosis and in congenital acanthocytosis which is a haemolytic anemia related to abnormal phospholipid metabolism [1]. This confirms our expected results.

*d) Presentation of patient 4 results:* Elliptocytes are elongated, oval, or ellipsoid red blood cells with rounded ends Fig. 10.

Table XI shows the colorimetric characteristics of elliptocytes. On analysis of the data, we find that the proportion of white pixels varies between 17% and 41%, indicating the presence of annulocytes. Analysis of the morphological data in Table XII shows a certain similarity between some forms of red blood cells and elliptocytes, as the compactness is between 0.9 and 1 and indicates the same range as that of healthy red blood cells and annulocytes. Indeed, compactness alone does not allow elliptocytes to be characterized. To solve this problem, we have calculated the interval r of the red blood cell count. This deviation between 25 and 30 pixels is greater, unlike the deviation on annulocytes, which is less than or equal to 7. So, the identification of elliptical shapes is precise by combining the compactness, and the distance r between the two axes. This combination allows our algorithm to accurately detect elliptocytes. In addition, we can look at the distribution of white and red pixels which indicates another form of anaemia. As a result, the patient may suffer from 2 or more forms of anemia.

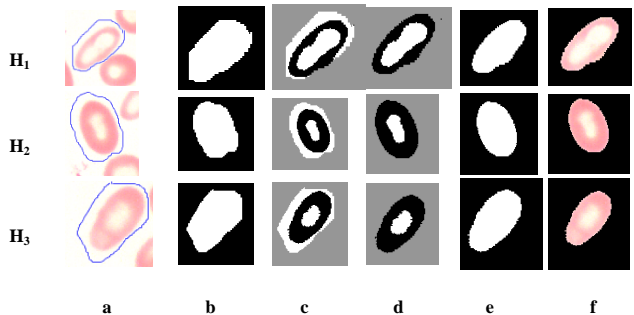

Fig. 10. (a) Region of Interest in the Source Image (b) Selection Area Binarized by the Otsu Algorithm (c) Isolation of the Erythrocyte (d) Segmentation into Two Classes (e) Isolated Segmented Erythrocyte (f) Isolated Source Erythrocyte.

TABLE XI. COLORIMETRIC PARAMETERS OF ELLIPTOCYTES

| | Class size | | Color | | | Area | Distribution | |
|---|---|---|---|---|---|---|---|---|
| | Red Pixels | White Pixels | Red | Green | Bblue | In pixel | % White | % Red |
| $H_1$ | 2297 | 1628 | 255 | 233 | 228 | 3925 | 41.45 | 58.55 |
| $H_2$ | 3637 | 771 | 255 | 233 | 228 | 4408 | 17.50 | 82.50 |
| $H_3$ | 3449 | 784 | 255 | 233 | 228 | 4233 | 18.52 | 81.48 |

TABLE XII. MORPHOLOGICAL PARAMETERS OF ELLIPTOCYTES

| | Compc | Area | Peri | MinoA | MajoA | spacing | varconvex |
|---|---|---|---|---|---|---|---|
| $H_1$ | 0,90 | 3925 | 236 | 23.29 | 52.64 | 29.36 | 0 |
| $H_2$ | 1.09 | 4408 | 225 | 28.51 | 53.51 | 25.00 | 0 |
| $H_3$ | 1,02 | 4233 | 228 | 25.00 | 51.60 | 26.60 | 0 |

## V. CONCLUSION

In this document, we have proposed a new method for identifying the different forms of red blood cells through color microscopic images of blood smears taken from anaemic patients. The proposed method is based on a selection of different regions of interest each containing a target red blood cell to which we apply the application developed in this work in order to extract morphological and colorimetric parameters for the different related red blood cells. Moreover, the application allows the precise identification of certain types of anemia. This formal identification allows pathologists to improve the diagnosis and establish a treatment protocol adapted to each type of anemia. This work, whose objective is to contribute to improving the well-being of our populations, cannot stop there because the list of pathologies, whether anemic or not related to the color and morphology of red blood cells, is still long. In the future, we will continue the detection of new forms of red blood cell disease not studied here in order to improve our semi-automatic diagnostic application.

Our future work concerns the following points:

- Development of algorithms to detect cells whose outline is not clear.
- Determine erythrocyte inclusions: Heinz bodies, Howell-Jolly bodies, etc.
- Determine a threshold at which the diagnosis of the type of anemia will be clearly made.
- Considerably reduce the rendering time of the results.
- Extend our method to other types of biological examinations that are still difficult to perform in our laboratories.

**Data Availability:** The figures and images used to support the results of this study are included in the article.

**Conflicts of Interest**: The authors declare no conflict of interest.

ACKNOWLEDGMENTS

We think the Blood Transfusion Center (BTC) and the Regional Hospital Center (RHC) of Yamoussoukro. The authors thank you for your sincere collaboration in the production of this paper: Dr D. Bosso (BTC of Yamoussoukro), Dr C. Houedanou (RHC of Yamoussoukro), K. Hermann (RHC of Yamoussoukro), Dr B. Dembélé (BTC of Abidjan), K. Eugene (RHC of Yamoussoukro), and T. Abalé (BTC of Abidjan).